\documentclass[%
superscriptaddress,
preprint,
amsmath,amssymb,
]{revtex4-2}
\usepackage{ulem}
\usepackage{xcolor}
\usepackage{multirow}
\usepackage{graphicx}
\usepackage{dcolumn}
\usepackage{bm}
\usepackage{hyperref}

\begin{document}


\title{Absence of magnetic order and magnetic fluctuations in RuO$_{2}$}

\author{Jiabin Song}
\affiliation{Beijing National Laboratory for Condensed Matter Physics and Institute of Physics, Chinese Academy of Sciences, Beijing 100190, China}
\affiliation{Collaborative Innovation Center for Nanomaterials Devices, College of Physics, Qingdao University, Qingdao 266071, China}

\author{Chao Mu}
\affiliation{Beijing National Laboratory for Condensed Matter Physics and Institute of Physics, Chinese Academy of Sciences, Beijing 100190, China}
\affiliation{School of Physical Sciences, University of Chinese Academy of Sciences, Beijing 100190, China}

\author{Shilin Zhu}
\affiliation{Beijing National Laboratory for Condensed Matter Physics and Institute of Physics, Chinese Academy of Sciences, Beijing 100190, China}

\author{Xuebo Zhou}
\affiliation{Beijing National Laboratory for Condensed Matter Physics and Institute of Physics, Chinese Academy of Sciences, Beijing 100190, China}

\author{Wei Wu}
\affiliation{Beijing National Laboratory for Condensed Matter Physics and Institute of Physics, Chinese Academy of Sciences, Beijing 100190, China}

\author{Yun-ze Long}
\affiliation{Collaborative Innovation Center for Nanomaterials Devices, College of Physics, Qingdao University, Qingdao 266071, China}

\author{Jianlin Luo}
\email{jlluo@iphy.ac.cn}
\affiliation{Beijing National Laboratory for Condensed Matter Physics and Institute of Physics, Chinese Academy of Sciences, Beijing 100190, China}
\affiliation{School of Physical Sciences, University of Chinese Academy of Sciences, Beijing 100190, China}

\author{Zheng Li}
\email{lizheng@iphy.ac.cn}
\affiliation{Beijing National Laboratory for Condensed Matter Physics and Institute of Physics, Chinese Academy of Sciences, Beijing 100190, China}
\affiliation{School of Physical Sciences, University of Chinese Academy of Sciences, Beijing 100190, China}


\begin{abstract}

A novel magnetic class blending ferromagnetism and antiferromagnetism, termed altermagnetism, has gained significant attention for its staggered order in coordinate and momentum spaces, time-reversal symmetry-breaking phenomena, and promising applications in spintronics. Ruthenium dioxide (RuO$_{2}$) has been considered a candidate material for altermagnetism, yet the presence of magnetic moments on Ru atoms remains a subject of debate. In this study, we systematically investigated the magnetic properties of RuO$_{2}$ powder using nuclear quadrupole resonance (NQR) measurements. The NQR spectra show that there is no internal magnetic field. Furthermore, the temperature independence of spin-lattice relaxation rate, $1/T_1T$, proves that there are no magnetic fluctuations. Our results unambiguously demonstrate that Ru atoms in RuO$_{2}$ possess neither static magnetic moments nor fluctuating magnetic moments, and thus RuO$_{2}$ does not possess the magnetic characteristics essential for altermagnetism.

\end{abstract}


\maketitle

\section{Introduction}
In the field of magnetic materials, the discovery and characterization of diverse magnetic phases have long been a focal point of intensive research. The concept of altermagnetism introduces a novel perspective in magnetism, integrating characteristics such as zero net magnetization similar to antiferromagnets and nonrelativistic spin splitting analogous to ferromagnets\cite{Smejkal2022a, Smejkal2022b}. These unique characteristics endow altermagnets with potential applications in spintronics\cite{Smejkal2022c}. Ruthenium dioxide, RuO$_{2}$, crystallizing in the $P4_2/mnm$ rutile structure with Ru$^{4 +}$ ions in a 4d$^4$ electron configuration, has been theoretically proposed as a potential altermagnetic system\cite{Berlijn2017, Ahn2019, Rafael2021, Smejkal2022a, Smejkal2023, McClarty2024}. However, RuO$_{2}$ has been considered an ordinary Pauli paramagnetic (i.e., nonmagnetic) metal from the perspective of electronic properties \cite{Ryden1970}.

Based on neutron diffraction and resonant x-ray scattering experiments \cite{Zhu2019,Berlijn2017}, it has been suggested that RuO$_{2}$ exhibits antiferromagnetic (AFM) ordering with a high N${\rm \acute{e}}$el temperature $(> 300$ K) and a Ru magnetic moment size of $\sim 0.05$ $\mu_{B}$. These claims have generated significant interest in its potential applications in spintronic devices\cite{Bose2022}. Strain-induced superconductivity further highlights the need for precise characterization of its electronic and magnetic properties\cite{Ruf2021}.

Following the inference of AFM ordering in RuO$_{2}$, various theoretical predictions and experimental results related to transport phenomena have been reported. Theoretical studies have predicted anomalous Hall effects linked to the collinear AFM phase and the noncentrosymmetric position of the nonmagnetic oxygen\cite{Rafael2021}, and experimental results supporting these predictions have been reported\cite{Feng2022}. Spin current due to the spin-splitter effect generated in the AFM phase has also been theoretically proposed, followed by reports of experimental results in favor of the prediction\cite{Shao2021}. The occurrence of the chirality magneto-optical effect has also been theoretically predicted\cite{Zhou2021}.

Notably, while altermagnet like behavior has been reported in RuO$_{2}$ film samples\cite{Zhu2019,Bai2022,Bai2023,Feng2022,Tschirner2023,Feng2024,Liao2024,Guo2024,Fedchenko2024,Li2025,Zhang2025}, bulk RuO$_{2}$ materials are more likely to show an absence of magnetic order\cite{Hiraishi2024,Liu2024,Philipp2024,Kiefer2025}. The reported size of Ru magnetic moments is close to the limit of sensitivity in neutron measurements\cite{Berlijn2017}. Additionally, the observations of symmetry forbidden reflections can be attributed to multiple diffraction\cite{Philipp2024,Kiefer2025}. Moreover, a recent theoretical study suggests that AFM ordering may be induced by hole doping due to Ru vacancies in RuO$_{2}$, which is intrinsically nonmagnetic\cite{Smolyanyuk2024}. Thus, there is a high demand for verification of the AFM phase with local magnetic probes that are complementary to diffraction experiments. Understanding the magnetic properties of bulk RuO$_{2}$ is not only crucial for fundamental research in condensed matter physics, but also has potential implications for applications in spintronics.

Nuclear magnetic resonance (NMR) is a powerful tool for detecting magnetic properties in altermagnets\cite{Jiang2025}. A previous $^{99}$Ru NMR study on RuO$_{2}$ demonstrated that the 4$d$ spin contribution is not seen\cite{Mukuda1999}. However, the $^{99}$Ru NMR study faced challenges due to the complex spectrum shape of $^{99}$Ru, which results from a large electric field gradient (EFG) and a large EFG asymmetry. Consequently, for the same reason, $^{101}$Ru spectra are too broad to be detected in NMR experiments\cite{Mukuda1999}. Moreover, NMR experiments need to apply a magnetic field which can melt the magnetic order\cite{Baek2017}. For example, $\alpha$-RuCl$_3$ possesses an ordered magnetic moment of $\sim 0.5$ $ \mu_{\rm B}$\cite{Banerjee2017}, which is ten times larger than that reported in RuO$_2$\cite{Berlijn2017}. Its magnetic order can be suppressed by an $8$ T magnetic field\cite{Zheng2017}. To overcome these limitations, we used nuclear quadrupole resonance (NQR) measurements in a zero magnetic field to study the magnetic and electric properties of RuO$_{2}$. Our results show that there is neither magnetic order nor magnetic fluctuations down to $3.3$ K, indicating that the Ru atoms in RuO$_{2}$ do not possess magnetic moments.

\section{Experimental details}
\begin{figure}
\includegraphics[width=0.45\textwidth,clip]{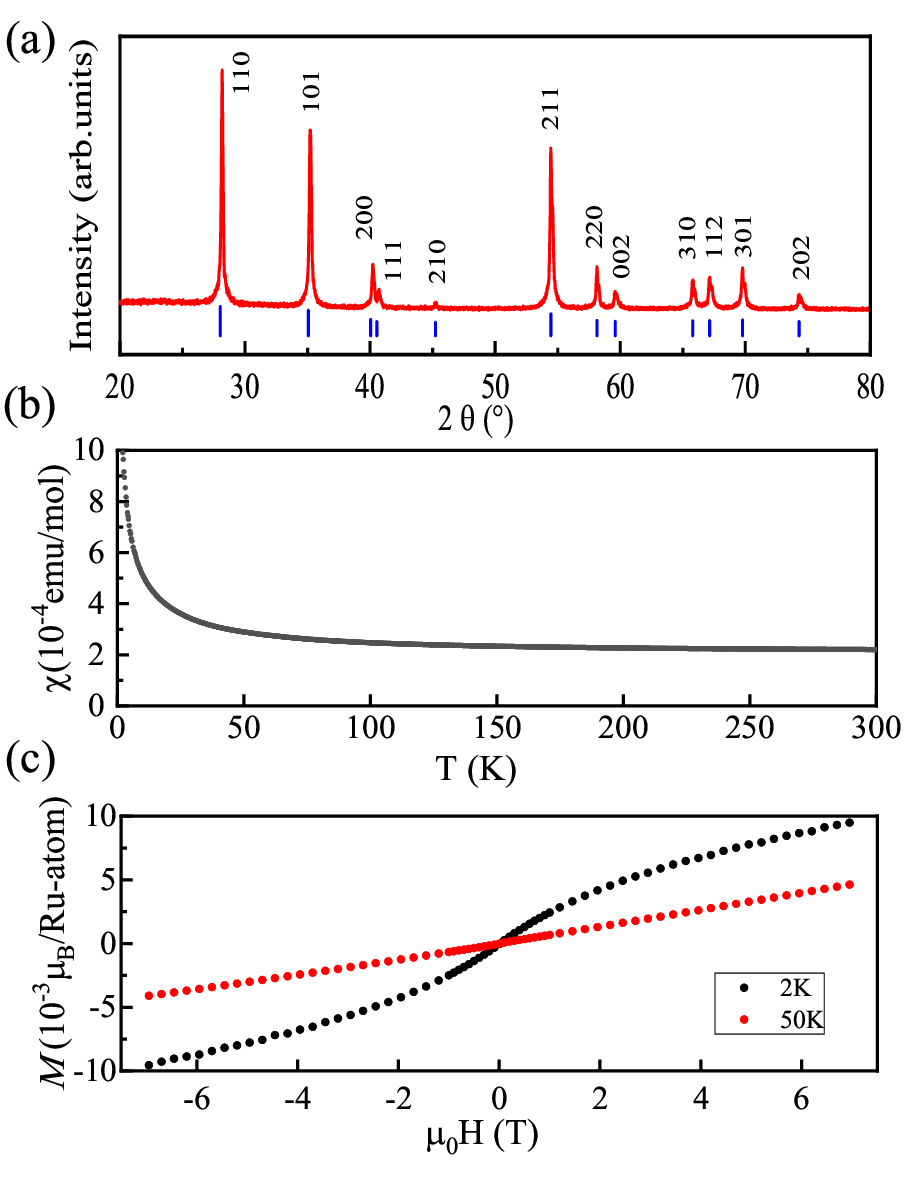}
\caption{\label{fig:xrd}(a) X-ray diffraction data obtained at room temperature for commercial RuO$_2$ powder. The pattern indicates a phase-pure sample without any detectable impurity peaks. (b) Temperature dependence of the magnetic susceptibility measured under a magnetic field of 0.5 T. (c) Magnetization as a function of magnetic field up to 7 T at temperatures of 2 K and 50 K.}
\end{figure}

The commercial RuO$_{2}$ powder with a purity of 99.95$\%$ was utilized in NQR experiments. The quality was ascertained by powder XRD, as shown in Fig. \ref{fig:xrd}(a). No impurity peaks were detected in the XRD pattern, confirming the high phase purity of the sample. Rietveld refinement was performed using a tetragonal structure model based on space group No. 136 ($P4_2/mnm$), yielding the lattice parameters
$a=4.486$(1) $\mathring{A}$ and $c=3.103$(1) $\mathring{A}$. These data are in the same range as the values reported in earlier studies\cite{xrd1969,xrd2024,Kiefer2025}. The magnetic susceptibility and magnetization results of the powder sample are shown in Figs. 1(b) and (c). Compared with single-crystal data, the susceptibility is larger over the entire temperature range attributed to paramagnetic moments\cite{Kiefer2025}.

The NQR measurements were performed with a phase-coherent spectrometer from Thamway Co. Ltd. The NQR spectra were acquired by integrating the intensity of spin echo at each frequency. To optimize the signal-to-noise ratio, the LC circuit was tuned and matched at each frequency. The spin-lattice relaxation time $T_{1}$ was measured at the $\pm 5/2 \rightarrow \pm 3/2$ peak using a single saturation pulse\cite{Narath1967}.

\section{Results and discussion}
Theoretically, the nuclear spin Hamiltonian of the interaction between the quadrupole moment $Q$ and the EFG can be written as\cite{Slichter1990}
\begin{equation} \label{eq:vQ} 
\begin{aligned}
\mathcal{H}_{\rm Q} = \frac{h\nu_{\rm Q}}{6}\left[ (3I_{z}^{2}-I^{2})+\frac{\eta}{2} (I_{+}^{2}+I_{-}^{2})  \right],
\end{aligned}
\end{equation}
where $\nu _{\rm Q}$ is the quadrupole resonance frequency along the principal axis, $h\nu _{\rm Q}  = 3eQV_{zz} /2I(2I - 1)$, and $h$ is the Planck constant. The asymmetry parameter of the EFG is defined as $\eta  = ( {V_{xx} - V_{yy} } )/V_{zz}$, where $V_{xx}, V_{yy}, V_{zz}$ are the components of the EFG along the local coordinates $x$, $y$, and $z$ directions respectively. In RuO$_2$, the local coordinates of the Ru atom are oriented along the [110], [1$\bar{1}$0], and [001] directions, respectively\cite{Ahn2019}. For nuclei with spin $I=5/2$, such as $^{99}$Ru and $^{101}$Ru, a zero-field NQR spectrum exhibits two transition peaks, namely $\pm 3/2\rightarrow\pm 1/2$ and $\pm 5/2\rightarrow\pm 3/2$.

\begin{figure}
\includegraphics[width=0.45\textwidth,clip]{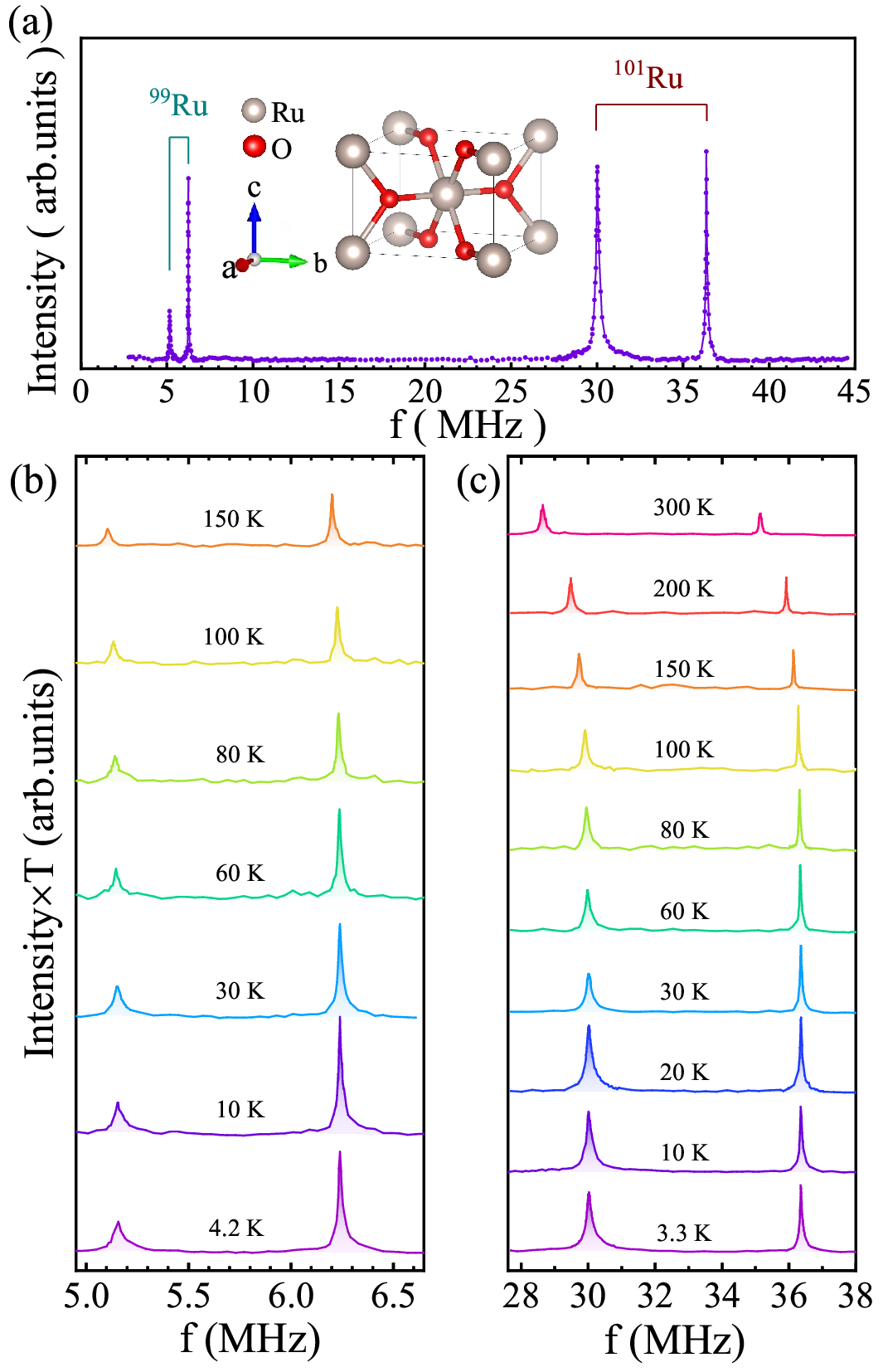}
\caption{\label{fig:nuQ}(a) The NQR spectrum of RuO$_2$ at $10$ K. The two low-frequency peaks are attributed to the $\pm 3/2\rightarrow\pm 1/2$ and $\pm 5/2 \rightarrow\pm 3/2$ transitions for $^{99}$Ru, while the two high-frequency peaks are attributed to $\pm 3/2\rightarrow\pm 1/2$ and $\pm 5/2 \rightarrow\pm 3/2$ transitions for $^{101}$Ru.
The NQR spectra at various temperatures of (b) $^{99}$Ru and (c) $^{101}$Ru. Baselines have been vertically offset for visual clarity.}
\end{figure}

Figure \ref{fig:nuQ}(a) shows the zero-field NQR spectrum of RuO$_2$ at $10$ K. As shown in the crystal structure illustration in Fig. \ref{fig:nuQ}(a), there is only a Ru site in RuO$_2$. The two low-frequency peaks originate from $^{99}$Ru, while the two high-frequency peaks originate from $^{101}$Ru. Their frequency ratio
$^{101}f/^{99}f = ^{101}Q/^{99}Q \sim 5.8$ indicates that there are only EFG contributions, without magnetic field contribution. If there is a significant internal magnetic field at the Ru site, each Ru isotope will show five resonance peaks, such as RuSr$_2$YCu$_2$O$_8$ and RuEu$_{1.4}$Ce$_{0.6}$Sr$_2$Cu$_2$O$_{10-\delta}$\cite{tokunaga2001,furukawa2003valence}, where the hyperfine field at the Ru site can reach up to $60$ T. If a small internal magnetic field exists, the NQR peaks should exhibit splitting. The amount of splitting $\Delta f $ is proportional to the internal field $H_{\rm int}$ and is given by the equation $\Delta f = 2 \gamma H_{\rm int}$, where $\gamma$ is the gyromagnetic ratio. If the internal field is not homogeneous, peaks will broaden and the broadening width is also proportional to the internal field. The narrowest peak of $\pm 5/2 \rightarrow\pm 3/2$ transitions for $^{99}$Ru has a full width at half maximum (FWHM) of $25$ kHz, limiting any potential internal field to less than $6.4$ mT. Using the inner-core polarization coupling constant of $22$ T/$\mu _{\rm B}$ reported in Ref. \cite{Mukuda1999}, the magnetic moment of Ru is constrained to be no larger than $3 \times 10^{-4}$ $\mu _{\rm B}$. The absence of splitting and broadening confirms the lack of an internal magnetic field and indicates that all Ru atoms occupy equivalent lattice sites.

The resonance peaks at various temperatures (ranging from 150 K to 4.2 K for $^{99}$Ru and 300 K to 3.3 K for $^{101}$Ru) are shown in Figs. \ref{fig:nuQ}(b) and (c). Notably, the $\pm 3/2\rightarrow\pm 1/2$ peaks exhibit a broader linewidth than the $\pm 5/2\rightarrow\pm 3/2$ peaks. This phenomenon arises because the $\pm 3/2\rightarrow\pm 1/2$ transition is more sensitive to $\eta$ dispersion than the $\pm 5/2 \rightarrow \pm 3/2$ transition when $\eta$ is large\cite{Chizhik2014}. As the temperature decreases, all peaks shift slightly toward higher frequencies. The lack of splitting and negligible broadening confirm the absence of a magnetic transition in bulk RuO$_{2}$ down to $3.3$ K, strongly suggesting the lack of magnetic order within this temperature range.

\begin{figure}
\includegraphics[width=0.45\textwidth,clip]{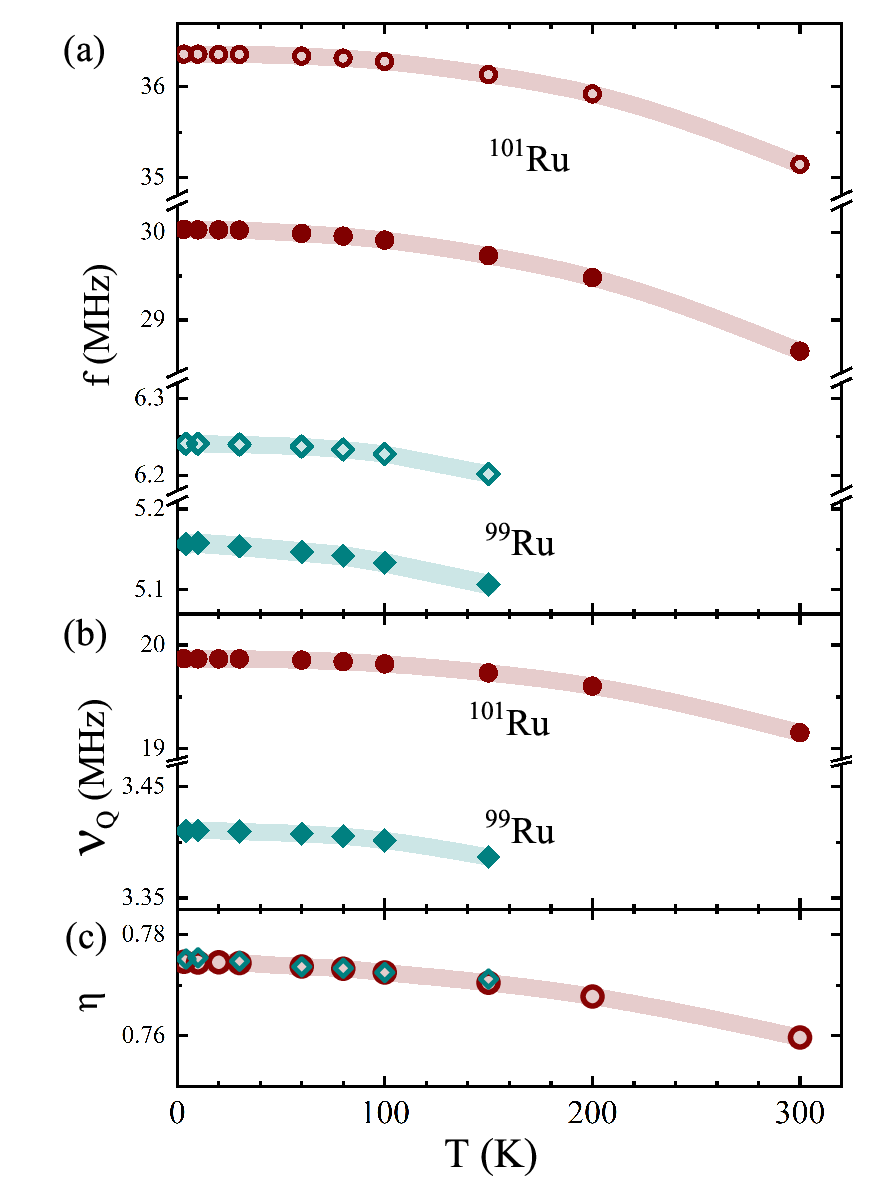}
\caption{\label{fig:eta}The temperature dependence of (a) peak positions of $\pm 3/2 \rightarrow \pm 1/2$ (solid) and $\pm 5/2 \rightarrow \pm 3/2$ (open) transitions, (b) quadrupole resonance frequency $\nu_\mathrm{Q}$, and (c) asymmetry parameter $\eta$ for $^{101}\mathrm{Ru}$ and $^{99}\mathrm{Ru}$, respectively.}
\end{figure}

We summarize the resonance frequencies of all peak positions in Fig. \ref{fig:eta}(a). The quadrupole resonance frequency $\nu_{\rm Q}$ and $\eta$ can be deduced by combining the pair peaks\cite{Chizhik2014}, shown in Figs. \ref{fig:eta}(b) and 3(c), respectively. $\nu _{\rm Q}$ and $\eta$ of both isotopes increase gradually with decreasing temperature. This phenomenon can be attributed to the shrinkage of the rutile structure as the temperature drops\cite{Sugiyama1991}. The contraction of the lattice leads to changes in the electronic environment around the Ru atoms, resulting in a slow increase in the quadrupole resonance frequency and the asymmetry parameters. The absence of sudden changes indicates that there is no phase transition. We notice that due to anisotropic position of the Ru atom, the site has a large $\eta$ value of $\sim 0.77$. Considering the relation $V_{xx}+V_{yy}+V_{zz}=0$, we can calculate that $V_{yy}/V_{xx}=7.7$. The directions of $V_{xx}, V_{yy}, V_{zz}$ at the central Ru atom are [110], [1$\bar{1}$0], and [001], while the directions at the apical Ru atom  are [$\bar{1}$10], [110], and [001]. The EFG of central Ru and apical Ru is rotated by $\pi/2$ along the crystallographic $c$ axis with respect to each other. So along the [110] or [1$\bar{1}$0] direction, EFG will alternate between the two values of $V_{xx}$ and $V_{yy}$. An electric field can exert a force on a magnetic dipole\cite{Goldhaber1989,Vaidman1990}, which may acquire a geometric phase and form Landau levels\cite{Nakata2017}. Moreover, there is moderate spin-orbit coupling (SOC) in RuO$_2$\cite{Occhialini2021}, which associates with a field-induced magnetic moment\cite{Song2025}. Since there is no magnetic order, this unique EFG and SOC may have a relationship with the spin-splitting effect in films\cite{Bai2022,Bai2023,Bose2022,Karube2022,Guo2024,Liao2024}.

\begin{figure}[htb]
\includegraphics[width=0.4\textwidth,clip]{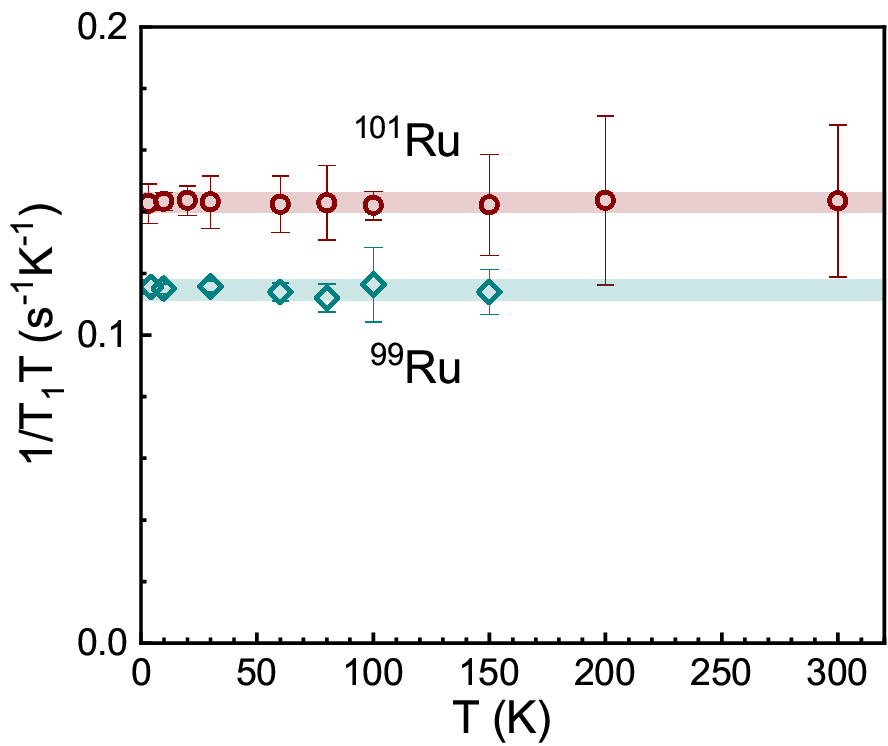}
\caption{\label{fig:T1T} Temperature dependence of $1/T_1T$ measured at the $\pm 5/2\rightarrow \pm 3/2$ peak for $^{101}$Ru and $^{99}$Ru, respectively. The lines are guides to the eye. The ratio of $^{101}T_{1}/^{99}T_{1} = (^{99}\gamma / ^{101}\gamma)^2 \sim 0.8$.}
\end{figure}

In order to check whether there are fluctuating magnetic moments of Ru, we measured the $T_{1}$ at $\pm 5/2 \rightarrow\pm 3/2$ peaks for both $^{101}$Ru and $^{99}$Ru. Due to the large $\eta \sim 0.77$, the relaxation function for spin $I = 5/2$ is $1-M(t)/M(\infty)=\frac{13}{44} {\rm exp}(-3.1t/T_{1})+\frac{31}{44} {\rm exp}(-8.2t/T_{1})$, where $M(t)$ is the nuclear magnetization at time $t$ after the saturation\cite{Chepin1991}. $1/T_{1}T$ describes how the nuclei arrive at their thermal equilibrium via the process of spin-lattice relaxation, and is proportional to the summation of the imaginary part of the dynamical susceptibility. Magnetic fluctuations with any momentum $q$, such as ferromagnetic fluctuations with $q=0$ or antiferromagnetic fluctuations with $q=\pi$, will enhance $1/T_{1}T$ when temperature decreases\cite{Li2012}. Conversely, if there is a transition to an ordered state, such as magnetic-order, charge order, and superconductivity, $1/T_{1}T$ should decrease below the transition temperature\cite{Li2012,Li2021}. As shown in Fig. \ref{fig:T1T}, for both $^{101}$Ru and $^{99}$Ru, $1/T_{1}T$ is temperature independent, similar to that in conventional metals. This demonstrates the absence of both magnetic fluctuations and a magnetic-order transition. Combined with the NQR spectra discussed above, we can conclude that the Ru atoms do not possess magnetic moments or magnetic multipoles. Moreover, the ratio of $^{101}T_{1}/^{99}T_{1}$, which is $(^{99}\gamma / ^{101}\gamma)^2 = 0.8$, indicates the absence of EFG fluctuations, despite the presence of a strong and asymmetric EFG. Therefore, RuO$_{2}$ is a conventional metal where electron-electron correlations are relatively weak.

\section{Summary}
In this study, NQR measurements were used to systematically probe the magnetic properties of bulk RuO$_2$. Our results unambiguously demonstrate that bulk RuO$_2$ exhibits neither magnetic order nor magnetic fluctuations of local magnetic moments. The large asymmetry parameter $\eta \sim 0.77$ reveals that along the [110] or [1$\bar{1}$0] direction the EFG at Ru sites alternates between the two values of $V_{xx}$ and $V_{yy}$, where $V_{yy}/V_{xx}=7.7$. Further investigation is required to determine whether this EFG behavior is associated with the spin-splitting effect in films.

\begin{acknowledgments}
This work was supported by the National Key Research and Developm ent Program of China (Grants No. 2022YFA1602800, No. 2022YFA1403903), the National Natural Science Foundation of China (Grants No. 12134018, No. 52325201),  and the Strategic Priority Research Program and Key Research Program of Frontier Sciences of the Chinese Academy of Sciences (Grant No. XDB33010100), and the Synergetic Extreme Condition User Facility (SECUF).

\end{acknowledgments}

\end{document}